\documentclass[epj,referee]{svjour}
\usepackage[dvips]{graphicx}

\def\be{\begin{equation}}
\def\te{\end{equation}}
\def\ee{\end{equation}}
\def\ba{\begin{eqnarray}}
\def\bea{\begin{eqnarray}}

\def\tea{\end{eqnarray}}
\def\ea{\end{eqnarray}}
\def\eea{\end{eqnarray}}

\begin{document}
\title{Kinesin and the Crooks fluctuation  theorem}
\author{Esteban Calzetta\inst{1}} 
%
%
\institute{Depto. de F\'{\i}sica, F. C. E. y N. - UBA and CONICET, Argentina}
\date{Received: date / Revised version: date}
%
\abstract{
The thermal efficiency of the kinesin cycle at stalling is presently a matter of some debate, with published predictions ranging  from $0$ (A. W. C. Lau, D. Lacoste and K. Mallick, Phys. Rev. Lett. 99, 158102 (2007); D. Lacoste, A. W. C. Lau and K. Mallick, Phys. Rev. E78, 011915 (2008)) to $100\%$ (G. Oster and H. Wang, in \emph{Molecular Motors}, edited by M. Schliwa (Wiley-VCH Verlag GmbH, Weinheim (2003), p. 207).  In this note we attemp to clarify the issues involved. We also find an upper bound on the kinesin efficieny by constructing an \emph{ideal kinesin cycle} to which the real cycle may be compared. The ideal cycle has a thermal efficiency of less than one, and the real one  is less efficient than the ideal one always, in compliance with Carnot's theorem.
} 
\maketitle
The application of physics thinking to life is one of the great breakthroughs of modern science \cite{Sch45}. The big questions (what is the fundamental difference between living and nonliving matter? is there life anywhere else in the Universe? \cite{Dar01}) are conspicuously evading us, but the progress in the analysis of concrete phenomena is astonishing \cite{Nel04,KusTur03}. Since this generally involves applying physical theory in contexts quite removed from the original ones, we physicists are learning quite a lot in the process.

The above said is particularly true of the application of thermodynamics to biophysical problems \cite{Hay08}. If a physical description of a living system is possible at all, it must include thermodynamics at some point. But when we focus on phenomena at the molecular level, it is the thermodynamics of small systems we are talking about, not the familiar macroscopic thermodynamics of heat engines and stars \cite{Rit07}.

The analysis of molecular motors is a case in point. Analyzing engines is what thermodynamics is good for; it was invented for just this purpose. But when we get to a particular molecular motor the possibility (let alone the usefulness) of a thermodynamic analysis seems somewhat less than obvious, and one correspondingly finds a variety of points of view and statements in the literature, even when addressing the same systems.

To be concrete, let us take the example of kinesin regarded as a molecular motor. Kinesin converts chemical energy into useful work; it should be possible to define the thermal efficiency as the ratio of output work to input energy. One expects this should be less than one by a finite amount, lest some kind of perpetual mobile could be built on kinesin. Indeed one can build a model of kinesin which predicts that the efficiency drops to $0$   at stalling \cite{LaLaMa07,LaLaMa08}. However a careful analysis reveals that the opposite conclusion of $100\%$ efficiency also has some grounds \cite{OstWan03}. 

Let us point out that the problem is not that there are several nonequivalent measures of efficiency going around \cite{LiDoZu05}. We are talking about thermal efficiency \cite{PBCB98,PaJiBr00}, and that is usually clearly distinguished from alternatives such as Stokes efficiency \cite{WanZho08}.

To understand how kinesin could reach the $100\%$ efficiency mark it is instructive to compare the kinesin cycle to any well known thermal engine, such as a gas engine going through an Otto cycle. The comparison is not arbitrary, since both kinesin and the gas engine extract their power from a chemical reaction, and are in contact with a single environment \cite{Mag94}. However, there is a potential difference between both machines. In the car engine, the fuel delivers energy to the working fluid under the form of heat. Therefore the entropy of the working fluid increases during combustion. Because of Clausius' inequality, the rest of the cycle cannot be completed adiabatically. To elliminate the surplus entropy, heat must be lost to the environment, and this detracts from the engine's efficiency. Note that this is true even under ideal conditions. After every form of friction is elliminated, efficiency remains less than one by a finite amount. Moreover, the ideal Otto is less efficient than the ideal Carnot cycle, if we set the hot and cold sources of the latter at the extreme temperatures of the former. 

In the case of kinesin, however, it is not obvious that heat is involved at any step in the process. Kinesin extracts its power from the free energy liberated by the hydrolysis of one ATP molecule. One can imagine that this free energy is delivered under the form of work and it is stored at intermediate steps as potential energy, for example, as elastic or electrostatic energy in the kinesin macromolecule \cite{CiSaTs06}. Work exchange is free from the restrictions imposed by the Second Law on heat exchange and heat to work conversion, and so a $100\%$ efficiency is not excluded.

Therefore the issue hinges on whether at any step within the kinesin cycle we may identify a conversion of work into heat, or at least an entropy increase. If this is the case, then the kinesin cycle cannot be completely adiabatic, and the efficiency must be strictly less than $100\%$. Moreover, this entropy increase must obtain even after known forms of friction, or of entropy production because of processes evolving at a finite rate, are elliminated. In other words, we ought to be able to identify an ideal kinesin cycle, and show that under no conditions the ideal kinesin cycle gets to $100\%$ efficiency. 

In a recent contribution, Bier \cite{Bie08} has presented a scenario along these lines. The kinesin cycle is an asymmetric hand on hand walk along a microtubule. At some point in the kinesin cycle, both kinesin heads are attached to the microtubule. Then one of the head detaches. According to \cite{Bie08}, while the head of kinesin is free it has many more available states than when docked, and so it also has a larger entropy. Therefore docking must be accompanied by an entropy decrease, and a corresponding heat loss $Q$ to the environment. The kinesin efficiency must therefore be less than $100\%$.

Our goal is to further advance this argument by estimating the free energy change $\Delta {\cal F}$ associated with a forward step of kinesin along the microtubule. Because kinesin is in contact with an environment at constant temperature, the free energy change poses an upper bound on the work $W$ which may be extracted from kinesin, $W\leq \left(-\Delta {\cal F}\right)$, with equality obtaining under ideal conditions.

The problem is how to extract $\Delta {\cal F}$ from actual experiments, since these are done under non ideal conditions. To overcome this difficulty we note that, if we regard a backward step as the time-reversal of a forward step, then $\Delta {\cal F}$ can be related to the ratio of the probabilities for forward and backward steps through the Crooks fluctuation theorem. Since these probabilities can be measured \cite{NHITY03,TNIY05,CarCro05}, this solves our problem. 

The rest of the paper is organized as follows. In the next section we provide some basic background material on the kinesin cycle. Then in the following section we present the Crooks fluctuation theorem and we argue that this theorem can be applied to the kinesin cycle. In Section IV we use it to estimate the free energy change and the efficiency of the process. We conclude with some very brief final remarks.

\section{A brief description of the kinesin cycle}
Let us begin by stating the essential facts about the kinesin cycle \cite{SVB00,How02,RCSNMVC03,Blo06}.

\begin{figure}[htp]
\centering
\includegraphics[scale=.4]{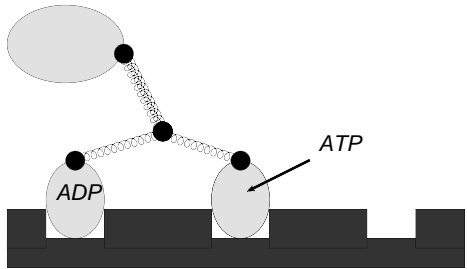}
\caption[first]{At the beginning of the kinesin cycle, both heads are attached to the microtubule. The rear head is associated to a molecule of adenosine diphosphate (ADP). The front head captures a molecule of adenosine triphosphate (ATP) from the environment. }\label{first}
\end{figure}

\begin{figure}[htp]
\centering
\includegraphics[scale=.4]{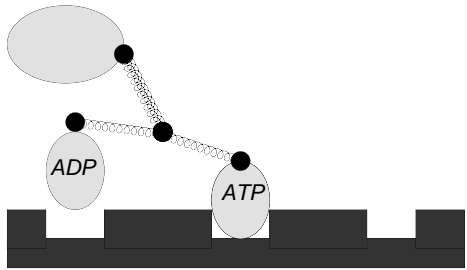}
\caption[second]{In the second step of the cycle, the rear head separates from the microtubule and the whole molecule swings around the front head, bringing the other head close to the next docking site. }\label{second}
\end{figure}

\begin{figure}[htp]
\centering
\includegraphics[scale=.4]{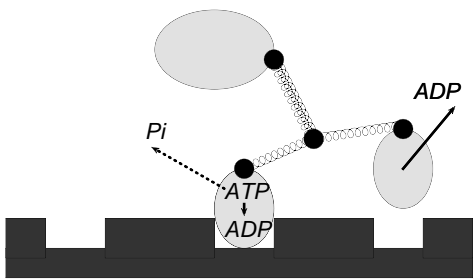}
\caption[third]{In the third step of the cycle, the now front head releases its ADP molecule into the environment, while in the other head ATP is decomposed into ADP and a phosphate (Pi), which is released. This process, namely the  hydrolysis of ATP, is the source of the energy for the whole cycle. }\label{third}
\end{figure}

\begin{figure}[htp]
\centering
\includegraphics[scale=.4]{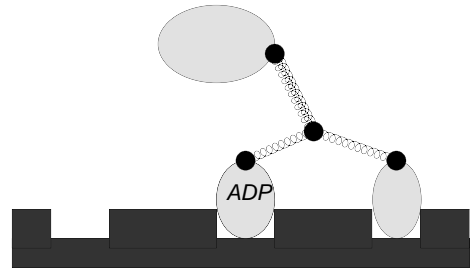}
\caption[fourth]{Finally, the front head binds to the microtubule. The cell disposes of the released ADP and Pi and provides a new ATP molecule. This is analogous to the exhaust stroke at the end of the Otto cycle. }\label{fourth}
\end{figure}

Kinesin is a polymer which can attach to a load on one end and has two heads on the other end. By attaching the heads to a microtubule, kinesin can perform an asymmetric hand on hand one-dimensional walk, effectively dragging the load against an external force $F$ and the viscous drag from the environment, thus doing useful work. Each step in this walk corresponds to a cycle. The strides are $\delta =8\;nm$ long, corresponding to the distance between the attachment sites on the microtubule. During one step one head remains firmly docked, while the other head traverses double that distance. We give in figures \ref{first} to \ref{fourth} a simplified account of the kinesin cycle.

One crucial input for the analysis below is the amount of energy liberated by the hydrolysis of one ATP molecule. We adopt the estimate provided by Bier \cite{Bie08}, namely
 $25\;k_BT$. Here $k_B=1.4\:10^{-23}\:J\:K^{-1}$ is the Boltzmann constant and $T=300\;K$ is the environment temperature. Let us observe that a lower value would change the numerical prediction for the efficiency, but not the essentials of the arguments below; only a value as low as half of this would validate the prediction of $100\%$ efficiency at stalling under ideal conditions. We assume that exactly one ATP molecule is decomposed in each cycle, so the number $r$ of steps per unit time is equal to the reaction rate for ATP hydrolysis. This reaction rate is regulated by the cell through the concentrations of the relevant substances. The dependence of $r$ with the load is weaker than other relevant factors \cite{CarCro05} and we shall disregard it.

Upon release, kinesin undergoes a configuration change which pushes the free head to the neighborhood of next unoccupied docking site. Kinesin can exert a constant force, mostly independent of the opposing external force. Otherwise, the free head is caught up in the thermal Brownian motion generated by the environment. Therefore the end of the cycle is undetermined: the free head can dock in the intended site, in which case the step is successfully completed, or else it may be dragged back to its initial position. Since the reaction rate is fairly constant, the ratio of success to failure is the main determinant of the overall progress rate. The probability of a successful forward step over that of failure has been measured to be \cite{CarCro05,Bie08}

\be
\frac{P_f}{P_b}=\exp\left\{\frac{\delta}{2k_BT}\left[F_{st}-F\right]\right\}
\label{drift}
\te
where $F_{st}=7\;pN$ is the stalling force. At higher loads, kinesin drifts backwards. 

It follows from eq. (\ref{drift}) that the maximum work kinesin can do against the external force is $F_{st}\delta =13.3\;k_BT$, which is about half of the available energy. This puts an absolute limit on thermal efficiency at about $50\%$ at any operating condition, given our previous estimate for the available energy.

In the real world, there are many sources of dissipation that help to explain the missing efficiency, such as viscous drag or friction related to the conformational changes in the polymer. However, in devising an \emph{ideal} cycle we are free to imagine that those sources of dissipation may be controlled. For example, we may dream of a \emph{reversible} kinesin cycle where the free head drifts so slowly that it is always in thermal equilibrium with the environment. In this case, the net viscous drag would be negligible. 

\section{The Crooks fluctuation theorem}
When the kinesin cycle is completed, the only changes are that the molecule has been displaced along the microtubule and work has been done against an external force, if any. Since the whole process is isothermal, the work $W$ extracted is bounded by the free energy drop when kinesin moves one step forward. The efficiency of the ideal kinesin cycle, when equality obtains, is therefore equal to the ratio of the free energy drop to the input energy.

To obtain the free energy change associated to one step from observable data, we shall apply the Crooks fluctuation theorem \cite{Cro98,Cro99,Cro00,Rit03,CRJSTB05,Kur05} to kinesin. Let us first make explicit our conventions. We call $Q$ the heat taken in by the system in a nonequilibrium process, and $W$ the work given out \emph{by the system}. The First Law is written as $\Delta U=Q-W$, where $U$ is the internal energy, and the Clausius inequality is $Q\leq T\Delta S$, where $T$ is the temperature and $S$ is the entropy. We define the free energy as ${\cal F}=U-TS$, so the Clausius inequality implies $W\leq-\Delta {\cal F}$ in any isothermal process. For present purposes, we do not need to discriminate between internal energy and enthalpy $H=U+PV$, and therefore we do not make a distinction between the (Helmholtz) free energy ${\cal F}$ and the Gibbs free energy ${\cal G}=H-TS$. 

The Crooks fluctuation theorem concerns a system in interaction with an environment at temperature $T$. The system depends on some external parameter $\lambda$. Initially the system is at state $A$. 
The parameter $\lambda$ is made to depend on time following a trajectory $\lambda_f$. We call $P_f$ the probability that the system ends up in state $B$, giving out work $W$. If now the evolution of $\lambda$ is reversed, we call $P_b$ the probability that the system, now starting from $B$, ends up in state $A$ giving out work $-W$. The Crooks fluctuation theorem states that

\be
\frac{P_f}{P_b}=\exp\left\{\frac{-1}{k_BT}\left[\Delta {\cal F} +W\right]\right\}
\label{Crooks}
\te
There are several proofs of the Crooks fluctuation theorem in the literature, each based on different assumptions on the microscopic dynamics of the system in question. Since we lack a microscopic model of kinesin, none of those proofs can settle the question of whether the Crooks fluctuation theorem can be applied to the kinesin cycle. 

A relevant observation is that the kinesin molecule is a very complex system but only a few of the variables defining its state (such as the position of the heads and whether they are associated to ADP and ATP molecules) are relevant to our discussion. While formal proofs of the Crooks fluctuation theorem usually assume that the initial and final states are equilibrium states, from the physical point of view we may expect that the theorem holds if the characteristic times of the process are long enough that  the unobservable part of the theory (the particular microscopic states of system and environment) is capable to adjust to the instantaneous value of the relevant macroscopic variables.  This only puts a mild bound on the characteristic times associated to the macroscopic evolution of the system. To make this bound precise we need an estimate for the relaxation times of the microscopic model, which we do not have at present.

This expectation is consistent with related work on fluctuation theorems applied to systems in nonequilibrium steady states \cite{Maes03,ZBCK05} and also on periodically driven systems \cite{BrHaSe04,BraSei04}.

The above argument only makes the wide applicability of Crooks fluctuation theorem plausible, but it is not conclusive by itself. In last analysis, we must consider whether the Crooks fluctuation theorem has been applied to situations comparable to the kinesin process.

Probably the most detailed match of the Crooks fluctuation theorem against experimental data is the one reported in \cite{CRJSTB05}. In this opportunity, the Crooks fluctuation theorem was used to estimate the free energy difference associated to the unfolding of a RNA molecule. In the experiment, a single molecule was repeatedly folded and unfolded. We find the periodic folding of a single molecule is analogous to the cycles of the kinesin displacement, and therefore the success of the application of the Crooks fluctuation theorem in one case argues for the feasibility of its application in the other.

\section{Kinesin and the Crooks fluctuation theorem}

To apply the Crooks fluctuation theorem to kinesin, we identify the macroscopic initial and final states $A$ and $B$ as the initial and final states in the kinesin cycle, and the external parameter $\lambda$ as any variable that measures the progress of the molecule from one pair of docking sites to the next, such as the expectation value of the position of the center of mass of the molecule along the microtubule. We also  regard a backward step as the time reversal of a forward step, both under the same external force $F$ (so that work is reversed). Noting that $W=F\delta$ for a forward step,  the Crooks fluctuation theorem \ref{Crooks} with the probability ratio given in \ref{drift} implies

\be
\Delta {\cal F}=\frac{-\delta}{2}\left[F_{st}+F\right]
\label{free}
\te
On the other hand, $-\Delta {\cal F}$ is the maximum work which may be extracted from the system at constant temperature. We have given above an estimate of the energy input as $25\;k_BT$. This is close to $2\delta F_{st}=26.6\;k_BT$\cite{Bie08}. Adopting this approximate energy input, the efficiency of the ideal kinesin cycle is

\be
\eta_{ideal}=\frac14\left[1+\frac F{F_{st}}\right]
\label{ideal}
\te
Under ideal circumstances, all the energy $2\delta F_{st}$ given to the polymer at the beginning of the cycle goes either into the reversible work $-\Delta {\cal F}$ or into the heat dissipated during docking. If docking is isothermal, then we have

\be
S_{free}-S_{dock}=\frac{Q_{dock}}{k_BT}=\frac{\delta}{2k_BT}\left[3F_{st}-F\right]
\label{deltas}
\te
where $S_{free}$ and $S_{dock}$ are the entropies of the free head and the docked states, respectively.

We see that the efficiency of the ideal kinesin cycle ranges from $25\%$ when gliding to $50\%$ at stalling (of course, the second  value will change if the assumed input energy is modified). No conflict with classical thermodynamics is apparent. The efficiency of the real cycle is less than this. There are several reasons for this reduced efficiency, of which two are outstanding. First, not all the available work is actually extracted: some of it is left to be dissipated as heat, for example by opposing the viscous drag or as excess kinetic energy to be absorbed by the docking site. Second, the cycle may fail, with the polymer drifting backwards rather than forward. Therefore the actual average work is

\be
\left\langle W\right\rangle =F\left\langle \delta\right\rangle
\label{mean}
\te
where $\left\langle \delta\right\rangle$ is the average displacement

\be
\left\langle \delta\right\rangle =\delta\tanh\left\{\frac{\delta}{4k_BT}\left[F_{st}-F\right]\right\}
\label{meand}
\te
and the actual efficiency is

\be
\eta_{real}=\frac{F\left\langle \delta\right\rangle}{2F_{st}\delta}
\label{real}
\te
Observe that $\eta_{real}$ vanishes at both gliding and stalling, in agreement with \cite{LaLaMa07,LaLaMa08}. In general, we have $\eta_{real}<\eta_{ideal}$ always, also in agreement with Carnot's theorem: no actual machine can perform better than a reversible machine with the same energy input.

\section{Final Remarks}
The issue of the efficiency of molecular motors, with kinesin as a particular case, is a matter of debate in the literature. If one regards the kinesin cycle as a series of conversions of one form of potential energy into another, always through the action of work, then the efficiency can be very high. In particular, at stalling, when no systematic viscous drag opposes motion, the efficiency could be as high as $100\%$.

On the other hand, if one adopts the phenomenological definition of efficiency eq. \ref{real}, then one can say that the efficiency is zero at stalling by definition.

In this note, we have attempted to clarify the issues involved by noting that 
for a motor like kinesin,  operating at constant temperature, the efficiency is limited by the free energy change associated to each step forward along the microtubule. 

The most important contribution of this note is the suggestion that this free energy change can be extracted from actual experimental data on the ratio of probabilities for a forward and a backward step, through the application of Crooks fluctuation theorem. Our actual analysis yields an efficiency of  $50\%$ at stalling. Therefore the prediction of $100\%$ efficiency ought to be discarded. On the other hand, our analysis concerns only an ideal situation, and therefore does not conflict with a prediction of $0\%$ under physical circunstances.

The central issue is therefore if the application of Crooks fluctuation theorem is warranted. It is possible to verify the Crooks fluctuation theorem for a given microscopic dynamical model, but not to actually prove its validity in an experimental situation. Very much the same could be said about the Second Law itself. In this case, support for the application of the Crooks fluctuation theorem to kinesin may be found by noting the extreme generality of this statement, and that it has already been applied in similar situations.

The ultimate motivation of our work is that the thermodynamic analysis provides guidance for further microscopic studies of kinesin as a dynamical system. From this point of view, we expect the final test of the hypothesis advanced in this note will be its heuristic value.

\section{Acknowledgments}

This work is supported in part by University of Buenos Aires, CONICET and ANPCyT (Argentina)

%
%

\end{document}